# A silicon photonic add-drop filter for quantum emitters

SHAHRIAR AGHAEIMEIBODI,[1] JE-HYUNG KIM,[2] CHANG-MIN LEE,[1] MUSTAFA ATABEY BUYUKKAYA,[1] CHRISTOPHER RICHARDSON,[3] EDO WAKS[1,4, *]

[1]*Department of Electrical and Computer Engineering and Institute for Research in Electronics and Applied Physics, University of Maryland, College Park, Maryland 20742, United States*
[2]*Department of Physics, Ulsan National Institute of Science and Technology, Ulsan 44919, South Korea*
[3]*Laboratory for Physical Sciences, University of Maryland, College Park, Maryland 20740, United States*
[4]*Joint Quantum Institute, University of Maryland and the National Institute of Standards and Technology, College Park, Maryland 20742, United States*
*\*edowaks@umd.edu*

**Abstract:**

Integration of single-photon sources and detectors to silicon-based photonics opens the possibility of complex circuits for quantum information processing. In this work, we demonstrate integration of quantum dots with a silicon photonic add-drop filter for on-chip filtering and routing of telecom single photons. A silicon microdisk resonator acts as a narrow filter that transfers the quantum dot emission and filters the background over a wide wavelength range. Moreover, by tuning the quantum dot emission wavelength over the resonance of the microdisk we can control the transmission of the emitted single photons to the drop and through channels of the add-drop filter. This result is a step toward the on-chip control of single photons using silicon photonics for applications in quantum information processing, such as linear optical quantum computation and boson sampling.

## 1. Introduction

Integrated photonics is a promising toolbox for realizing quantum information processors that are capable of optical quantum computation [1–3], quantum simulation [4,5], and quantum communication [6,7]. Single photons are excellent carriers of quantum information and can be generated efficiently using quantum emitters [8]. Major advances have been made to integrate such high-quality single photon sources with low loss silicon-based optical waveguides. For example, several works have demonstrated the integration of solid-state quantum emitters with $SiO_2$ [9], SiN [10–14], and Si photonic chips [15,16]. Moreover, recent advances in developing on-chip single-photon detectors [17–19] have paved the way for quantum photonic circuits that are fully chip-integrated (i.e., the photons do not leave the chip from generation to detection). Integrating these elements on the same chip enables compact quantum devices with robust and low loss operation as well as programmable functionality.

In addition to the generation and detection of photons, researchers have demonstrated different approaches for on-chip routing of single photons in the near infrared regime. The electro-optic effect [20] and optomechanical forces [21] of GaAs have been used to demonstrate routing of single photons in suspended structures containing quantum dots. Implementing single photon routing using silicon-based photonics rather than suspended GaAs waveguides could enable easier scaling because of the mature foundry-level fabrication available for this platform. A hybrid platform based on SiN waveguides and quantum dots hosted in GaAs has been used to implement routing and multiplexing of single photons in the near infrared regime [14]. However, SiN does not support fast reconfiguration mechanisms, which limits the ability to control the active filters in this platform to slow thermo-optic [14] or strain [22] tuning of the refractive index. In contrast, silicon photonics offers fast tuning by modulation of the carrier concentration using electrical contacts [23,24] or by introduction of field-induced second order nonlinearity [25]. To date, compact devices for routing and filtering quantum emitters with silicon photonics has yet to be implemented.

In this work, we demonstrate a silicon photonic add-drop filter that can filter and route telecom single photons from quantum dots. We use InAs quantum dots embedded in an InP nanobeam as bright sources of single-photon emission at the telecom band [26,27]. This emission efficiently couples to a silicon waveguide using an adiabatic taper. The filter is comprised of an ultracompact silicon microdisk resonator that is evanescently coupled to waveguides [28]. To experimentally implement this hybrid device, we used a pick-and-place technique based on scanning electron microscopy and focused ion beam milling to transfer the nanobeam containing the quantum dots to the silicon substrate [15,29]. With this design, we demonstrate on-chip filtering of the unwanted background emission over a wide wavelength range while the desired quantum dot emission is efficiently transmitted through the filter. Moreover, by temperature tuning the system we can control the emission wavelength of a quantum dot with respect to that of the resonator and demonstrate routing of the single photons between the drop and through channels. Filtering and routing single-photon emission in silicon photonics is an important step toward fully integrated quantum photonic circuits that generate, process, and detect light on a compact chip.

## 2. Device design and working principle

Figure 1a shows the general working principle of our proposed hybrid device, which features an InP nanobeam embedded with InAs quantum dots as well as a silicon microdisk resonator. Single photons from the quantum dot couple to the transverse electric mode of the InP nanobeam. A Bragg reflector (period of 290 nm and radius of 100 nm) on one end of the nanobeam directs the photons in a single direction. An adiabatic taper on the other side of the InP nanobeam efficiently transfers the single photons to the silicon waveguide, which is also tapered to match the optical mode of the nanobeam. Both tapers in the coupling region are 5 μm long.

The add-drop filter consists of a silicon microdisk resonator and two waveguides (drop and through channels) evanescently coupled at the top and bottom of the microdisk. Telecom single photons that are transferred to the silicon waveguide are routed to the drop channel if resonant with the microdisk. Otherwise, they propagate to the through channel of the add-drop filter. By tuning the wavelength of either the quantum dot or the microdisk resonator we can essentially switch the single photons between the drop and through channels.

Figure 1b shows the simulated propagation of light at the intersection of the InP nanobeam and the silicon waveguide when they are placed side by side. Using a finite-difference time-domain simulation (Lumerical, FDTD), we calculated a coupling efficiency exceeding 95% at 1300 nm, which is the typical emission wavelength of our quantum dots. We performed variational finite-difference time-domain simulations (Lumerical, Mode) to design the resonance wavelength and linewidth of the resonator. Considering the linewidth of our quantum dots (~0.1 nm), a quality factor larger than 1000 for the microdisk resonator is not desirable for achieving efficient transfer of the photons to the drop channel due to losses from filtering the quantum dot emission spectrum. To maintain a small footprint, we used a disk radius of 2.125 μm, gap of 90 nm between the drop/through channels and microdisk, and a waveguide width and height of 230 nm and 220 nm, respectively. From simulation using these parameters, we obtained a resonance of around 1300 nm with a quality factor of 620. Figure 1c and 1d show the simulated transmission of light to the through and drop channels, respectively. The silicon waveguide width at the curves is 400 nm to avoid additional bending losses. The width tapers down to 230 nm to match the optical mode of the disk resonator over a length of 5 μm. Both drop and through channels end with periodic grating couplers (550 nm period and 50% duty cycle) with a 37% outcoupling efficiency for the transverse electric mode.

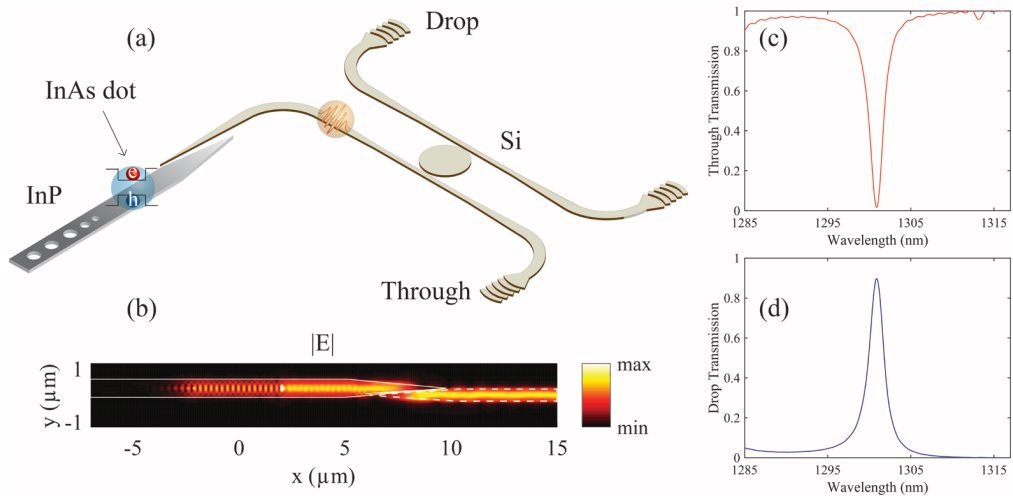

Figure 1. (a) Illustration of the hybrid device containing InAs quantum dots embedded in an InP nanobeam and a silicon photonic microdisk resonator. (b) Simulated light propagation in the coupling region. The colormap represents the electric field intensity (|E|). White solid and dashed lines demonstrate the boundary of the InP nanobeam and silicon waveguide, respectively. (c,d) Simulated transmission of light to the (c) through and (d) drop channels.

## 3. Device fabrication and measurement methods

We fabricated our hybrid device based on a pick-and-place approach [15,29] to integrate the InP nanobeam and the silicon add-drop filter, which were prepared individually on separate chips. First, we patterned the nanobeam using electron beam lithography on a substrate containing 280 nm thick InP on a 2 μm thick AlInAs sacrificial layer. We then used dry and wet etching to transfer the pattern from the resist to the InP membrane and suspend the structures, respectively. We chose a 220 nm thick silicon-on-insulator substrate to define our add-drop filter and used electron beam lithography to pattern the resist. Then we deposited Cr as a metal mask. Finally, we etched the substrate with inductively coupled plasma etching and removed the Cr. To transfer the nanobeam from the host substrate to the silicon-on-insulator substrate, we attached a microprobe to the nanobeam and used focused ion beam etching to remove the supporting tethers. Figure 2a shows a false color scanning electron microscopy image of the integrated device. Figure 2b is a further magnified view of the coupling region where the InP and silicon tapered waveguides are placed side-by-side.

To characterize the routing and filtering performance of the device, we used a micro-photoluminescence setup operated at 4 K. We excited the quantum dots using a 780 nm continuous wave laser focused down by an objective lens (NA = 0.7) to a diffraction-limited spot. We collected the photoluminescence signal from the quantum dots through the same objective lens and sent the collected photoluminescence to a grating spectrometer (Princeton Instruments) equipped with a nitrogen-cooled InGaAs detector array.

To characterize the response of the microdisk resonator, we used the broad emission of the ensemble quantum dots, which had an inhomogeneous broadening of more than 200 nm. Therefore, at high excitation powers (orders of magnitude higher than the saturation power of the quantum dots) they act as integrated broadband light sources. By collecting the photoluminescence signal from the drop and through channels and normalizing by the total signal, we obtained the response of the microdisk resonator. Figures 2c,d show the transmission spectra of the through and drop channels respectively, when we excited the nanobeam with a power of 520 µW. We observe multiple dips (peaks) in the spectrum of the through (drop) channel that are separated by a free spectral range of ~ 16 nm.

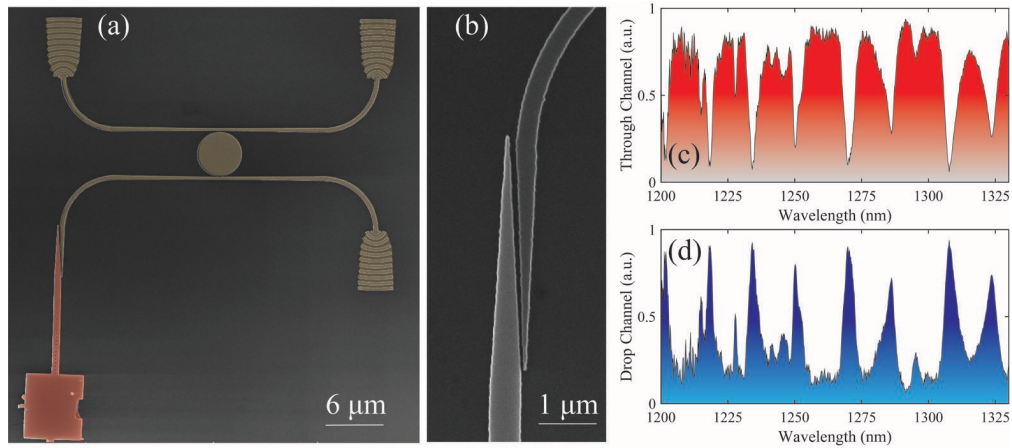

Figure 2. (a) False color scanning electron microscopy image of the fabricated hybrid device. Red and yellow represent InP and silicon, respectively. (b) A magnified view of the adiabatic tapering in the coupling region. The nanoscale accuracy of our pick-and-place technique allows alignment of these two sub-micron structures. (c,d) Measured normalized intensity at the (c) through and (d) drop channels when exciting the ensemble of the quantum dots above their saturation power.

## 4. Results

To isolate the single quantum dot emission, we lowered the excitation power to 5 µW which is below the dot saturation level. Figure 3a shows the photoluminescence spectrum when we excited the quantum dots and collected the signal directly from the nanobeam. We observed multiple quantum dot emission lines which is consistent with our laser spot size and the quantum dot density of $10\ \mu m^{-2}$. Next, we measured the photoluminescence spectrum from the drop channel while still exciting the quantum dots with the same power (Figure 3b). We observed a significant suppression of the photoluminescence signal while a narrow wavelength range at around 1234 nm transferred to the drop channel. The quantum dot emission highlighted by the red shaded box (Figures 3a and 3b) is on resonance with the microdisk resonator (Figure 3c) and therefore efficiently transferred to the drop channel while other emission lines were strongly suppressed. There is also a small signal observed at the drop channel around 1250 nm (Figure 3b), which may be a quantum dot emission coupled to another resonator mode. Moreover, multiple quantum dot emission lines couple to the resonator modes with wavelengths over 1300 nm (blue shaded box). The background suppression range can be expanded by designing resonators with a free spectral range larger than the inhomogeneous broadening of the quantum dots.

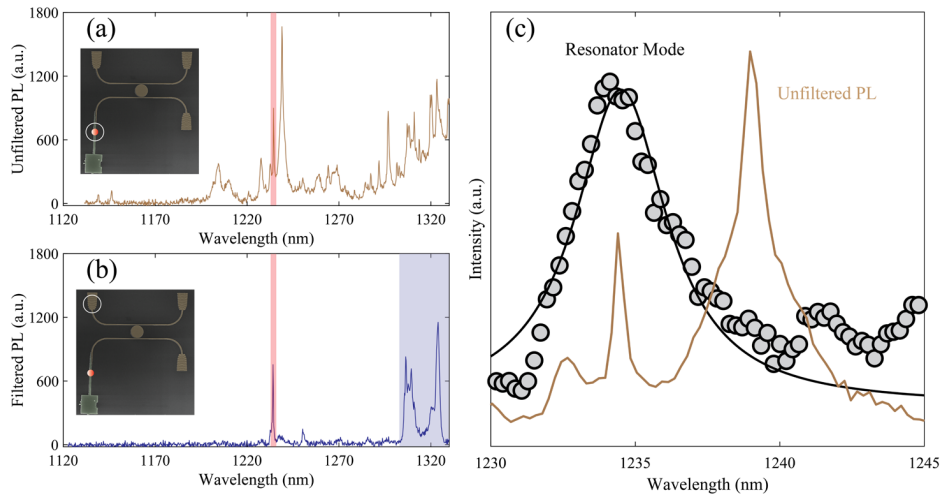

Figure 3. (a) Photoluminescence spectrum of the quantum dots when the excitation and collection spots are located on the nanobeam. (b) Photoluminescence spectrum of the quantum dots when the excitation and collection spots are located on the nanobeam and drop channel, respectively. The insets show the excitation and collection schemes, where the orange dot and the white circle are the excitation and collection spots, respectively. The shaded red boxes show the quantum dot emission that is transferred to the drop channel. (c) Same spectrum as (a) overlapped with the resonator mode. Black dots are the measured data and the solid curve is a Lorentzian fit to the data.

Next, we show that emission of a quantum dot can be routed to either the drop or through channels by tuning the dot wavelength with respect to the resonance of the microdisk. We used temperature tuning as a simple tool to tune the dot on and off resonance with the microdisk resonator. To be able to temperature-tune the emission wavelength of a quantum dot over the linewidth of the resonator we used one of the resonator modes (centered around 1364 nm) that has the highest quality factor ($Q = 661$ obtained with a Lorentzian fit) and a quantum dot that is slightly blue-shifted from the resonator mode at 5 K. The shaded gray curve in Figure 4a shows a Lorentzian fit to the microdisk resonator response at the drop channel. The dashed vertical lines represent the emission wavelength the quantum dot at different temperatures. By varying the temperature of the sample from 5 K to 50 K, we tuned the emission wavelength over the entire resonance mode of the device. We note that we did not observe a shift in the resonance wavelength of the microdisk resonator in our measurements. This is because the refractive index of silicon is insensitive to temperature at cryogenic temperatures [30].

We collected the photoluminescence signal of the dot from both the through and drop channels at different temperatures. Figure 4b shows the normalized integrated intensity at different detunings (obtained at different temperatures) from the microdisk resonance for both the through and drop channels. We observed that when the dot is resonant with the resonator, it propagates to both the drop and through channels. However, when the emission wavelength is detuned from the resonance of the microdisk it mostly propagates to the through channel. We obtained a linewidth of $1.7\ nm$ from Lorentzian fits to the integrated intensity of Figure 4b, which is consistent with the $2.06 \pm 0.4\ nm$ linewidth from the gray curve in Figure 4a. Ideally, all the emission should transfer to the drop channel at the resonance wavelength. However, fabrication imperfections in our silicon device prevents critical coupling between the waveguide and the resonator and leads to a small extinction ratio (i.e., the ratio between the maximum and minimum transmission) of 1.65.

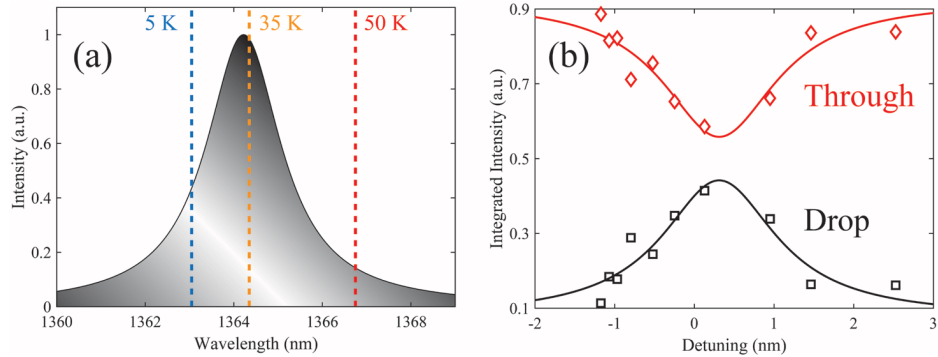

Figure 4. (a) A Lorentzian fit to the microdisk resonance (gray area) and center wavelength of the selected quantum dot at 5 K (blue), 35 K (orange), and 50 K (red). We tuned the temperature of the sample to tune the dot on- or off-resonance with the resonator. (b) The normalized integrated intensity of the dot for different detunings measured at the drop channel (black) and through channel (red).

## 5. Conclusion

In summary, we have implemented a hybrid platform containing a silicon microdisk resonator and InAs quantum dots embedded in an InP nanobeam for filtering single photons at telecom wavelengths. Using our integrated device, we achieved filtering of the background emission over a wide wavelength range while the desired quantum dot emission transfers through the filter. Moreover, by temperature-tuning a quantum dot over the resonance mode of the microdisk we controlled the transmission of the emission to the drop and through channels. Improving the design of the filter for a higher free spectral range can enable suppression of the background over the entire range of the ensemble quantum dot emission. Furthermore, better design and fabrication accuracy is required to increase the extinction ratio of the routing. Incorporating electrical contacts for electro-optic switching of the router can achieve GHz modulation of the single photons. With improved extinction ratios and ultrafast switching available in silicon photonics, this platform may find applications in single-photon de-multiplexing [31], realizing linear optical quantum computing [32,33], and boson sampling [34].

## Funding